\documentclass[letterpaper, 10 pt, conference]{ieeeconf}  

\IEEEoverridecommandlockouts                              

\usepackage{graphics} 
\usepackage{epsfig} 
\usepackage{amsmath} 
\usepackage{amssymb}  

\usepackage{caption}
\usepackage{subcaption}
\usepackage{xcolor}
\usepackage{multirow}

\title{\LARGE \bf
Inverter Output Impedance Estimation in Power Networks: A Variable Direction Forgetting Recursive-Least-Square Algorithm Based Approach  
}

\author{Jaesang Park\textsuperscript{1,a}, Alireza Askarian\textsuperscript{1,b}, and Srinivasa Salapaka\textsuperscript{1,c}
\thanks{\textsuperscript{1} Department of Mechanical Science and Engineering, University of Illinois at Urbana-Champaign, 61801 IL, USA
\textsuperscript{a}jaesang4@illinois.edu,\textsuperscript{b}askaria2@illinois.edu,\textsuperscript{c}salapaka@illinois.edu}%
}

\begin{document}

\maketitle
\thispagestyle{empty}
\pagestyle{empty}

\newcommand{\vasucommentx}[1]{{\bf \color{red} [XXX #1 XXX] }}
\newcommand{\jsresponse}[1]{{\bf \color{blue} [XXX #1 XXX]}}

\begin{abstract}
As inverter-based loads and energy sources become increasingly prevalent, accurate estimation of line impedance between inverters and the grid is essential for optimizing performance and enhancing control strategies. This paper presents a non-invasive method for estimating output-line impedance using measurements local to the inverter. It provides a specific method for signal conditioning of signals measured at the inverter, which makes the measured data better suited to estimation algorithms. An algorithm based on the Variable Direction Forgetting Recursive Least Squares (VDF-RLS) method is introduced, which leverages these conditioned signals for precise impedance estimation. The signal conditioning process transforms measurements into the direct-quadrature (dq) coordinate frame, where the rotating frame frequency is determined to facilitate a simpler and more accurate estimation.  This frequency is implemented using a secondary Phase-Locked Loop (PLL) to attenuate grid voltage measurement variations.   By isolating the variation-sensitive q-axis and relying solely on the less sensitive d-axis, the method further minimizes the impact of variations. The VDF-RLS estimation method achieves rapid adaptation while ensuring stability in the absence of persistent excitation by selectively discarding outdated data during updates. Proposed conditioning and estimation methods are non-invasive; estimations are solely done using measured outputs, and no signal is injected into the power network. Simulation results demonstrate a significant improvement in impedance estimation stability, particularly in low-excitation conditions, where the VDF-RLS method achieves more than three time lower error compared to existing approaches such as constant forgetting RLS and the Kalman filter.
\end{abstract}

\section{INTRODUCTION}
The rapid expansion of distributed energy resources (DERs) and inverter-based loads has made inverter-based grids increasingly common, driving the need for precise power regulation and deeper insights into grid interactions. In this context, output line impedance—the impedance between the inverter and the grid—plays a crucial role in determining inverter performance, affecting power injection limits and droop control characteristics \cite{hossen2020self, zhong2016universal}. Improved impedance estimation, as highlighted in \cite{10381844}, enhances controller bandwidth, while impedance also serves as an indicator of grid stiffness and assists in islanding detection \cite{jarraya2019online}. This capability supports smooth operational transitions for inverters \cite{10644914}. As inverter-based resources continue to expand, accurate monitoring of these factors is essential for maintaining grid stability.

\begin{figure}[tbh]
    \centering
        \begin{subfigure}[b]{0.5\columnwidth}
        \includegraphics[scale=0.35]{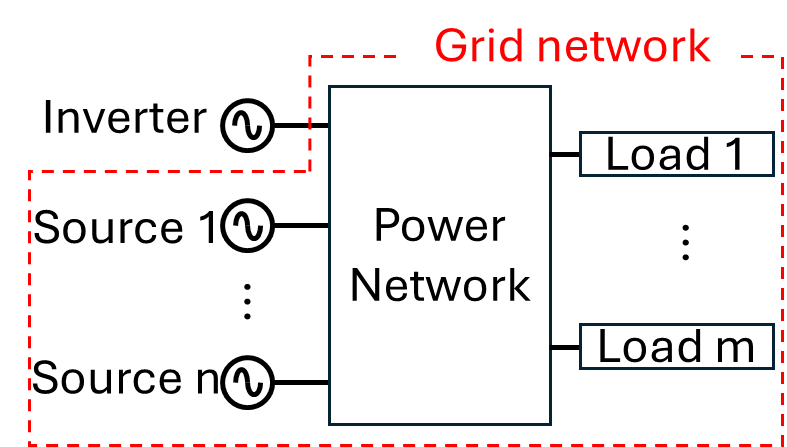}
        \caption{}
    \end{subfigure}%
    \begin{subfigure}[b]{0.5\columnwidth}
        \includegraphics[scale=0.35]{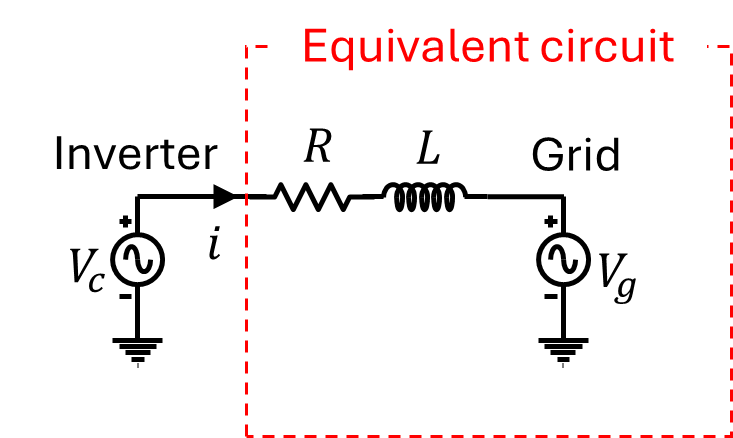}
        \caption{}
    \end{subfigure}\\
    \caption{Grid model: (a) Grid with a complex structure, and (b) Thevenin equivalent model from the inverter, where R and L represent the equivalent resistance and inductance of the Thevenin equivalent model.}
    \label{fig:circuits}
\end{figure} 
Output impedance is commonly defined as the impedance between the inverter and the grid, where the grid, in this context, represents an abstraction of the remaining network. In a complex power system, we model the grid as perceived by the inverter using Thevenin’s theorem as an equivalent grid voltage source in series with the output line impedance (see Fig. \ref{fig:circuits}). This Thevenin-equivalent impedance can vary significantly due to changes in the power network, such as fluctuations in electrical loads, the addition or removal of power sources, and environmental factors like temperature. Consequently, real-time impedance estimation is crucial for optimizing inverter performance and reliability. By continuously adapting to dynamic grid conditions, it enables stable power injection and ensures that control strategies remain effective.

The main challenges in accurately estimating output line impedance stem from several factors. (i) First, inverters typically lack access to global measurements or network-wide data, which makes it difficult to estimate the effective grid voltage. (ii) Additionally, measured signals often lack the necessary persistence of excitation, which is crucial for accurate impedance estimation. (iii) Since inverters usually operate at a steady state, only local output voltage and current are measurable, while both line impedance and grid voltage influence these measurements, making it essential to distinguish between their effects. (iv) Finally, in most grids, altering the power system to assist in impedance estimation is either impractical or not allowed. As a result, the estimation must be non-invasive, relying solely on locally measured signals without injecting real power disturbances into the system.  

To address challenges (i) and (ii), many line impedance estimation methods rely on signal injection. In \cite{ciobotaru2007line}, a high-frequency signal is injected assuming that the grid voltage contains only the nominal frequency component, helping to separate line impedance from grid voltage. However, this can distort the output voltage, and tampering with power flow is often not allowed. In \cite{ghanem2017grid}, the pulse-width modulation (PWM) signal used by inverters is leveraged to avoid additional distortion. However, it requires fast sensing beyond the PWM switching frequency, and filters on inverters reduce the injected signal, resulting in a low signal-to-noise ratio. In \cite{vasquez2009adaptive}, a time-domain differentiation method perturbs the inverter's power, assuming constant grid voltage, with the resulting voltage and current changes attributed to line impedance. However, injection-based methods, along with concerns about voltage quality and power control, are often impractical due to grid regulations or operational constraints.

Another important class of methods addresses challenges (i), (ii), and (iv) (lack of persistent excitation while avoiding signal injection) by utilizing historical current and voltage data to estimate line impedance. These techniques leverage variations in the inverter's operating point over time. For instance, \cite{cobreces2007complex} assumes constant grid voltage in the direct-quadrature $(d,q)$ frame and uses a Recursive Least Squares (RLS) algorithm to estimate both line impedance and grid voltage. However, this algorithm continuously processes all data, meaning that changes in line parameters only have a noticeable impact when sustained over a long period to overcome the influence of previous data. As a result, updating the estimation to reflect new values becomes a slow process.

To speed up convergence, \cite{jarraya2019online} introduces the Constant Forgetting Recursive Least Squares (CF-RLS) method, which discounts older data using a forgetting factor.  While this method enhances the adaptation speed to changing parameters, the exponential discounting of historical data can lead to a loss of relevant information, which is especially significant in the absence of persistent excitation.  This issue is particularly important since many inverters operate in grid-following (GFL) mode, where they track a fixed current set point that seldom changes. Thus, even aggregated historical data lacks richness for effective impedance estimation. In \cite{fang2020grid}, line impedance is modeled as a dynamic state in a time-varying system defined using measured inverter voltage and current. A Kalman filter is then implemented to estimate the output line impedance. This algorithm demonstrates performance similar to the CF-RLS algorithm, sharing the same issue of gradually losing relevant information in the absence of persistent excitation.

In this work, we address challenges (i)-(iv) and also address the gaps of the above methods. We address these challenges in two steps -We first condition the measurement signals for better estimation before feeding them into the line parameter estimation algorithm. The conditioning step ensures our assumptions on grid attributes are practical and simplify the estimation process. Our approach operates in the $d$-$q$ coordinate frame, using a rotating frame frequency tied to the inverter rather than the grid.  To generate rotating frame frequency, we design a {\em secondary Phase Locked Loop (PLL)}, which is distinct from the usual PLL used for inverter control and droop regulation.  This secondary PLL facilitates a coordinate frame where there is a frequency separation between current signals and grid voltage dynamics, by leveraging the algebraic structure of the constitutive equation relating the inverter and grid voltages. In this equation, the line parameters appear as coefficients of measured current signal while the uncertainty due to grid voltage comes as an additive disturbance. Thus frequency separation simplifies the estimation problem. Unlike the inverter's primary PLL, which has a fast bandwidth to quickly track frequency changes, the secondary PLL has a lower bandwidth provides which provides better frequency separation between grid frequency changes and its phase difference variations. 
Additionally, we demonstrate that while the $d$-axis dynamics are less sensitive to phase difference, the $q$-axis dynamics are highly sensitive. Therefore, by focusing on the less sensitive $d$-axis dynamics, we effectively reduce fluctuations in the estimation.

In the second step, to address the issue of non-persistent excitation, we propose the use of the Variable Direction Forgetting Recursive Least Squares (VDF-RLS) method \cite{9143247} for line impedance estimation. Similar to RLS and CF-RLS, VDF-RLS manages historical data using an information matrix. However, when updating the matrix with new data, VDF-RLS compares the direction of the new information vector with the singular vectors of the previous information matrix. The forgetting factor is then applied selectively, discounting only the singular values corresponding to directions aligned with the new information vector. We demonstrate that this method can estimate changing line parameters by aggregating information efficiently over time while remaining robust to noise, even when measures signals have low signal-to-noise ratio.

Upon implementation, we demonstrate that the proposed VDF-RLS-based algorithm not only tracks parameter changes rapidly when excitation is present but also remains stable, achieving significantly lower error—up to three times smaller—even in the absence of excitation. This effectively balances adaptation speed and stability. Additionally, we show that the proposed preconditioning methods efficiently reduce estimation error by mitigating measurement noise and noise induced by inverter activity.

\section{Grid Modeling and Preconditioning for Impedance Estimation}
\subsection{Thevenin-Based Grid Representation}

The grid around an inverter is simplified using Thevenin's theorem as a single voltage source and output line impedance. The stiff voltage source, unaffected by the inverter’s operation, is treated as the grid voltage, while the impedance represents the output line impedance perceived by the inverter. For simplicity, we  approximate the Thevenin impedance with a first-order model consisting of resistance $R$ and inductance $L$. The dynamics that describe this system in Fig. \ref{fig:circuits} are given by 
\begin{align}
    \begin{split}\label{eqn:line_dym_general}
        \overrightarrow{V_c} = R \overrightarrow{i} +L \frac{d \overrightarrow{i} }{dt} + \vec{V_g},
    \end{split}
\end{align}
where $\overrightarrow{V_c}$, $\overrightarrow{V_g}$, $\overrightarrow{i}$, and $R$ and $L$ respectively represent  phasors of the voltage across the inverter's output capacitance, the grid voltage, the inverter current, and the line impedance parameters.

If we represent the magnitudes of the grid voltage, the inverter voltage and the output current respectively by $\overline{V_g}$, $\overline{V_c}$, $\overline{i}$, their phasors in a stationary frame can be expressed as:
\begin{align}\label{eqn:phasors_st}
\begin{split}
        \overrightarrow{V_g}(t)&=\overline{V_g}(t)e^{j(\int_0^t \omega_{grid}(\tau)d\tau+\phi_{grid}) }\\
        \overrightarrow{V_c}(t)&=\overline{V_c}(t)e^{j(\int_0^t \omega_{inv}(\tau)d\tau+\phi_{inv}) }\\
        \overrightarrow{i}(t)&=\overline{i}(t)e^{j(\int_0^t \omega_{curr}(\tau)d\tau+\phi_{curr})},  
\end{split}
\end{align}
where $\omega_{grid}$, $\omega_{inv}$, and $\omega_{curr}$ represent the frequency of each signal, ensuring $\phi_{grid}$, $\phi_{inv}$, and $\phi_{curr}$ remain constant.

Also, (\ref{eqn:phasors_st}) can be rewritten using the rotating frequency $\omega$ as:
\begin{align}
\begin{split}\label{eqn:phasors_rt}
        \overrightarrow{V_g}(t)&=\overline{V_g}(t)e^{j(\int_0^t \omega(\tau)d\tau+\psi_{grid}(t)) }\\ 
        \overrightarrow{V_c}(t)&=\overline{V_c}(t)e^{j(\int_0^t \omega(\tau)d\tau+\psi_{inv}(t)) }\\
        \overrightarrow{i}(t)&=\overline{i}(t)e^{j(\int_0^t \omega(\tau)d\tau+\psi_{curr}(t)) },    
\end{split}
\end{align}
where  $\psi_{grid}$, $\psi_{inv}$, $\psi_{curr}$ represent the phases of each signal in the rotating frame. The direct-quadrature $($d$-$q$)$ components $z_d$ and $z_q$ of any phasor $\vec z$ in the rotating frame are given by $z_d+jz_q:=e^{-j\int_0^t\omega(\tau)d\tau}\vec z$. By applying this coordinate transformation, equations (\ref{eqn:phasors_rt}) and (\ref{eqn:line_dym_general}) can be rewritten as:  
\begin{align}
    \begin{split}\label{eqn:line_dym_general_dq}
        V_c^d &= R i^d+ L\frac{d i^d}{dt}-\omega L i^q +V_g^d\\
        V_c^q &= R i^q+ L\frac{d i^q}{dt}+\omega L i^d +V_g^q,
    \end{split}
\end{align}
where $V_c^d=\overline{V_c}\cos(\psi_{inv})$, $V_c^q=\overline{V_c}\sin(\psi_{inv})$, $V_g^d=\overline{V_g}\cos(\psi_{grid})$, $V_g^q=\overline{V_g}\sin(\psi_{grid})$. Here, the goal is to estimate the parameters $R$ and $L$, given the measured signals $V_c(t)$ and $i(t)$. Here $V_g(t)$ is assumed to be not known, while $\overline{V_g}$ is assumed to be near constant. 

Estimating $R$ and $L$ is more accurate when there is a clear frequency separation between $V_g^{d,q}$ and $i^{d,q}$ in (\ref{eqn:line_dym_general_dq}). In this case, high-frequency variations in $V_c^d$ (or $V_c^q$) used for estimation are less affected by the unmeasured $V_g^{d,q}$, allowing better parameter estimation from measured signals. However, if $V_g^{d,q}$ and the measured signals (e.g., $V_c^{d,q}$) share similar frequency components, estimation errors can arise. For instance, an incorrect yet valid estimation satisfying (\ref{eqn:line_dym_general_dq}) could be $R, L = 0$ and $V_g^{d,q} = V_c^{d,q}$.

Most of the literature assumes inherent frequency separation, often treating $V_g^{d,q}$ as constant, based on the premise that $\overline{V_g}$ and $\omega_g$ change slowly due to the grid's inertia. Consequently, many methods estimate constant impedance and grid voltage from measurements. 

\subsection{Preconditioning Strategy for Improved Estimation}
\subsubsection{Decoupling d-Axis and q-Axis Dynamics for Robust Estimation}
In this section, we relax the assumption that $V_g^{d,q}$ remains constant by analyzing its dynamics. Consider the approximation of $V_g^{d,q}$ around its nominal phase, $\psi_{grid,nom}$ as follows:
\begin{align}\label{eqn:vgdq_perturbation}
\begin{split}
   V_g^d &= \overline{V_g}\cos(\psi_{grid,nom}+\delta \psi_{grid})\\
    &\approx \overline{V_g}\big(\cos(\psi_{grid,nom})-\sin(\psi_{grid,nom})\delta\psi_{grid}\big)
\end{split}\\
\begin{split}
    V_g^q &= \overline{V_g}\sin(\psi_{grid,nom}+\delta \psi_{grid})\\
    &\approx \overline{V_g}\big(\sin(\psi_{grid,nom})+\cos(\psi_{grid,nom})\delta\psi_{grid}\big).
\end{split}
\end{align}
Here, the high-frequency component of $V_g^{d,q}$ stems from the terms $\sin(\psi_{grid,nom})\delta\psi_{grid}$ and $\cos(\psi_{grid,nom})\delta\psi_{grid}$ respectively.
We first consider reducing $\sin(\psi_{grid,nom})$ and $\cos(\psi_{grid,nom})$. Since both cannot be minimized simultaneously, we focus on the $d$-axis dynamics, reducing $\sin(\psi_{grid,nom})$ by keeping $\psi_{grid,nom}$ small. During normal grid operation, $\delta = \psi_{grid} - \psi_{inv}$ is typically small, so ensuring a small $\psi_{inv}$ helps keep $\psi_{grid}$ small. This is achieved through the PLL, which regulates $\omega(t)$ by adjusting $V_c^q(t)$ to maintain $V_c^q = \overline{V_c} \sin(\psi_{inv})$ small.

\subsubsection{Design of a Rotating Reference Frame}

Second, we aim to attenuate high-frequency components of  $\delta\psi_{grid}$ and, consequently, $\psi_{grid}$. From (\ref{eqn:phasors_st}) and (\ref{eqn:phasors_rt}), it is clear that $\psi_{grid}$ in the rotating frame is highly dependent on the chosen frequency $\omega$. Specifically, comparing (\ref{eqn:phasors_st}) and (\ref{eqn:phasors_rt}), we obtain:
\begin{align}\label{eqn:dpsi_dt} \begin{split} \psi_{grid}(t)=\int_0^t \big(\omega_{grid}(\tau)- \omega(\tau)\big)d\tau+\phi_{grid}. \end{split} \end{align}
This equation explicitly shows how $\omega$ can be adjusted to control $\psi_{grid}$, which can be achieved through a (secondary) PLL design. Noting that for $\phi_{grid}=0$, $V_c^q= \overline{V_c}\sin(\psi_{inv})\approx \overline{V_c}\psi_{inv}=\overline{V_c}\int \big(\omega_{grid}-\omega(\tau)\big) d\tau$ ; then the PLL dynamics, as described in \cite{yazdani2010voltage}, are given by:
\begin{eqnarray} &&\hat{\omega} = H(s) \hat{V_c^q} = H(s) \overline{V_c} \frac{1}{s} (\hat{\omega_{grid}} - \hat{\omega}) \nonumber\\ \label{eqn:vqtoomega}
&\Rightarrow& \hat \omega =\left(\frac{H(s)\overline{V_c}}{s+H(s)\overline{V_c}}\right){\omega}_{grid}:=Q(s){\omega}_{grid}\\\ \label{eqn:vqtpsi}
&\Rightarrow& \psi_{grid} =\frac{1}{s}\left(1-Q(s)\right)\omega_{grid}\\ \nonumber &&=\frac{1}{s}\left(\frac{1}{s+H(s)\overline{V_c}}\right){\omega}_{grid}
\end{eqnarray}
where $H(s)$ is a design transfer function. We design $H(s)$ as a low-bandwidth transfer function, which ensures that $\frac{1}{s}\left(1-Q(s)\right)$ has low bandwidth and therefore $\psi_{grid}$ has only low-frequency components as desired. 

{\em Remark:} In primary PLL, which is designed to assist the frequency regulation, the goal is to track $\omega_{grid}$;  $H(s)$ in that case is designed to have enough bandwidth so that $Q(s)$ in (\ref{eqn:vqtoomega}) has bandwidth large enough to react to high-frequency variations in $\omega_{grid}$.

{\em Remark:} This preconditioning is essential since, as discussed, $\psi_{grid}$ dynamics correlate with $V_c^q$ and ultimately with $i^{d,q}$.
If the inverter attempts to follow a new setpoint by adjusting $i^{d,q}$, it will induce a change in $\delta$, which will ultimately lead to a change in $\psi_{grid}$.
With high bandwidth PLL, perturbations in $\psi_{grid}$ remain significant and cannot be treated as uncorrelated noise in estimation algorithms, leading to substantial estimation errors.

\subsubsection{Reformulating the Estimation Problem with Preconditioning}
We reformulate the estimation problem that exploits the preconditioning step, where we have obtained signal separation between the unmeasurable grid voltage component $V_g^{d}$ and the other measured signals. 
Some studies use the difference between two consecutive time instances, which is equivalent to taking a discrete-time derivative. This approach amplifies high-frequency noise and fails to effectively reject it, even when a low-pass filter is applied.
Thus we can use a band-pass filter (BPF) that attenuates both $V_g^{d}$, which is dominant only in low frequencies, and very high-frequency noise, while retaining moderately high-frequency components of measured signals $i_d, i_q$ and $d i_d/dt$. Accordingly we modify   (\ref{eqn:line_dym_general_dq}) to reformulate the relevant constitutive dynamics for the estimation problem as
{\small
\begin{align}
    \begin{split}\label{eqn:line_dym_general_delta}
        \underbrace{BPF(V_c^d(t))}_{y}&=\underbrace{\begin{bmatrix}BPF(i^d(t))\\\big(\frac{BPF({di^d(t)/dt})}{\omega_0}-\left(\frac{\omega(t)}{\omega_0}\right) BPF(i^q(t))\big)\end{bmatrix}^\top}_{u^\top}\underbrace{\begin{bmatrix}R\\\omega_0L\end{bmatrix}}_{\theta}\\
        &\quad+w.
    \end{split}
\end{align}
}
where $BPF(\cdot)$ corresponds to a taking band-pass filter and $w$ is noise.

Here we have used $\omega_0L$ instead of $L$ for better numerical conditioning since $BPF( i^d(t)$ and $\frac{\omega(t)}{\omega_0} BPF(i^q(t))$ are of the same order.  Without this, $BPF( i^d(t))\ll \omega(t) BPF(i^q(t))$ since $\omega \approx 377rad/s\gg 1$, which can cause numerical instability.  Once $\omega_0 L$ is estimated, $L$ can be easily obtained.  Also, in  (\ref{eqn:line_dym_general_delta}), we approximate $BPF(\omega(t) i^{d,q})$ by $\omega(t) BPF(i^{d,q})$ since $\omega(t)$ varies slowly under our secondary PLL preconditioning, making its effect under BPF negligible.  
Another advantage of the BPF operation is that $BPF(\frac{di^d}{dt})$, which inherently includes a low-pass filter component, avoids numerical issues with differentiating noisy $i^d$. 
In this paper, we use a band-pass filter of the form
\begin{align}
\label{eqn:bpf_eqn}
    BPF(s) = \left(\frac{\omega_1}{s+\omega_1}\right)\left(\frac{s}{s+\omega_2}\right)
\end{align}
where $\omega_1$ and $\omega_2$ represent the upper and lower cutoff frequencies, respectively.

The lower cutoff frequency is chosen to be 10 times the gain crossover frequency of $\frac{1}{s}H(s)$, while the upper cutoff frequency is selected to be an order of magnitude lower (i.e., at least ten times lower) than the control frequency. This ensures stable numerical computation of $d{i^d}/dt$, as discussed earlier.

\section{Estimation of line parameters: Recursive Least Squares estimation algorithms}

\subsection{RLS and CF-RLS}
We introduce the RLS and CF-RLS algorithms, previously used in line-parameter estimation \cite{cobreces2007complex, jarraya2019online}, as the foundation for our VDF-RLS method. To highlight their limitations and how VDF-RLS addresses them, we first outline these algorithms. In RLS and CF-RLS, the optimal estimate $\widehat{\theta}_K$ at each measurement instant $K$ is obtained by minimizing:
{\small\begin{align} J(\widehat{\theta},K)= \sum_{k=1}^K \lambda ^{K-k}\left(y_K-u^\top_k\widehat{\theta}\right)^\top \left(y_k-u^\top_k\widehat{\theta}\right),\end{align}}
where $\lambda\in(0,1]$ is the forgetting factor. In RLS based estimation ($\lambda=1$), all past data from the beginning of the estimation process is weighted equally, causing new measurements to have diminishing influence on  $\hat \theta_K$ as $K$ increases. Only measurements that persist over many time instances affect $\hat\theta$, only when the trend stays for many time instances. Therefore  RLS based estimation has less variation but is slow to adapt to changing parameters. In CF-RLS, $0<\lambda<1$, recent data is given higher weight, allowing faster adaptation. A smaller $\lambda$  
prioritizes new data, improving responsiveness to parameter changes, while a larger $\lambda$ reduces noise sensitivity, and decreases fluctuations in the estimation. Thus, CF-RLS balances adaptability and fluctuation-reduction by adjusting $\lambda$.

The optimal $\widehat{\theta}_K$ obtained from $\frac{\partial J}{\partial \widehat{\theta}}=0$ is given by:
{\small \begin{align}\label{eqn} \widehat{\theta}_K &= \bigg(\sum_{k=1}^K \lambda^{K-k}u_k u^\top_k \bigg)^{-1}\bigg(\sum_{k=1}^K\lambda^{K-k} u_k y_k\bigg). \end{align}}
This can be expressed in recursive form as: \begin{align}\label{eqn:CF-RLS}
\begin{split}
     \hat{\theta}_K     &=  \hat{\theta}_{K-1}+R_K^{-1}u_K\big(y_K-u^\top_K\hat{\theta}_{K-1} \big).\\
\end{split}
\end{align}
Here, $R_K$ is the information matrix, which can be updated recursively as: {\small \begin{align}
\label{eqn:CF-RLS_RKupdate}
        R_K=\bigg(\sum_{k=1}^K \lambda^{K-k}u_k u^\top_k\bigg)  = (\lambda R_k+u_K u^\top_K).
\end{align}}
With this recursive form, there is no need to store all the data; only the information matrix and the previous estimate are required to update the new estimate when a new measurement is obtained.

Additionally, by using the Woodbury matrix identity, the terms
$\hat{\theta}_k$ and $R_K$ in CF-RLS model (\ref{eqn:CF-RLS},\ref{eqn:CF-RLS_RKupdate}) can be expressed as:
\begin{align}\label{eqn:RLS_Kalman_form1}
    \begin{split}
    \hat{\theta}_K     &=  \hat{\theta}_{K-1} 
    +\frac{R_{K-1}^{-1}u_K}{\lambda+u^\top_KR_{K-1}^{-1}u_K}\big(y_K-u^\top_K\hat{\theta}_{K-1} \big)
    \end{split}
\end{align}
and
\begin{align}\label{eqn:RLS_Kalman_form2}
    \begin{split}
    R_K^{-1}&=\frac{1}{\lambda}R_{K-1}^{-1}-\frac{1}{\lambda}\frac{R_{K-1}^{-1}u_Ku^\top_KR_{K-1}^{-1}}{\lambda+u'\top_KR_{K-1}^{-1}u_K}\\
    &=R_{K-1}^{-1}-\frac{R_{K-1}^{-1}u_Ku^\top_KR_{K-1}^{-1}}{\lambda+u^\top_KR_{K-1}^{-1}u_K}\\
    &\quad+\left(\frac{1}{\lambda}-1\right)\bigg(R_{K-1}^{-1}-\frac{R_{K-1}^{-1}u_Ku^\top_KR_{K-1}^{-1}}{\lambda+u^\top_KR_{K-1}^{-1}u_K}\bigg).
    \end{split}
\end{align}

The main challenge in estimation is the lack of persistently rich data, leading to non-invertible or ill-conditioned information matrices. Specifically, for a fixed tolerance $\epsilon>0$, the information matrix effectively incorporates data from the past $K = \frac{\log\epsilon}{\log \lambda}$ measurements, as $\lambda^K \leq \epsilon$ beyond this time horizon. If $u(k)$ lacks sufficient new information within this period, $R_K$ becomes non-invertible or ill-conditioned. As seen in equation (\ref{eqn:CF-RLS_RKupdate}), this approach discards the entire previous information matrix, regardless of data quality.

In the context of power systems, it is common for the data to lack sufficient richness within such short intervals of time. For example, an inverter might maintain a constant current, meaning $u \approx 0$ over long periods, with only occasional set-point changes. In such scenarios, there is insufficient data richness for the information matrix to remain invertible within the time intervals dictated by a reasonable forgetting factor $\lambda$. In fact, during extended periods of constant current, the information matrix exponentially converges to the zero matrix.
More generally, with CF-RLS, if there is no persistent excitation—where the $2 \times 1$ input vector $u(t)$ lies along a single direction within the effective time horizon—the smaller singular value of $R_K$ will exponentially decay to zero at a rate determined by $\lambda$, except in the direction of $u$. Consequently, the singular value of $R_K^{-1}$ grows exponentially, making the estimation highly sensitive to that particular direction, as $R_K^{-1}$ dictates the update of the estimate, as seen in equation (\ref{eqn:CF-RLS}).
\subsection{Kalman Filter}
Another approach is to use a Kalman filter. In this method, the line impedance parameters are treated as state variables, and a time-varying system is formulated using measurements as follows:
\begin{align}
    \begin{split}\label{eqn:Kalman_statespace}
        \theta_{k+1} = I_2\theta_{k}+w_{p,k}\\
        y_k = u_k^\top \theta_k +w_{m,k},
    \end{split}
\end{align}
where $I_2$ is a $2\times 2$ identity matrix, and $\theta_k$, $y_k$, $u_k$ are defined in (\ref{eqn:line_dym_general_delta}). The terms $w_{m,k}$ and $w_{p,k}$ represent measurement noise and process noise, respectively.

The recursive update equations for Kalman filter-based parameter estimation are given by:
\begin{align}
\begin{split}\label{eqn:Kalman_recursion}
    &\hat{\theta}_{k+1} = \hat{\theta}_{k}+ P_ku_k(S_k+u_k^\top P_ku_k)^{-1}(y_k-u_k^\top\hat{\theta}_k)\\
    &P_{k+1} = P_k+Q_k-P_ku_k(S_k+u_k^\top P_ku_k)^{-1}u_k^\top P_k,
\end{split}
\end{align}
where $P_k$ represents a priori estimate covariance.

With the optimal choice of $S_k$ and $Q_k$, we set $S_k=\mathbb{E}[w_{m,k}^2]$ and $Q_k=\mathbb{E}[w_{p,k} w_{p,k}^\top]$, which correspond to the covariance of measurement noise and process noise, respectively. However, these values are typically treated as tunable parameters.

By comparing (\ref{eqn:Kalman_recursion}) with (\ref{eqn:RLS_Kalman_form1}) and (\ref{eqn:RLS_Kalman_form2}), it is evident that if we choose $S_k = \lambda$ and $Q_k = (\frac{1}{\lambda}-1)\left(R_{K-1}^{-1}-\frac{R_{K-1}^{-1}u_Ku^\top_KR_{K-1}^{-1}}{\lambda+u^\top_KR_{K-1}^{-1}u_K}\right)$, the Kalman filter and CF-RLS become identical, with $P_K=R_K^{-1}$. Notably, increasing $Q_k$ is implicitly equivalent to reducing $\lambda$ smaller, which increases the adaptation speed to new data.

Due to its structural similarity, the Kalman filter faces similar issues as CF-RLS when there is no persistent excitation. With the proposed conditioning method, when the inverter maintains constant voltage and current, its DC component becomes zero, leading to $u_k=0$. In such a scenario, the recursion of $P_k$ in (\ref{eqn:Kalman_recursion}) simplifies to $P_{k+1} = P_{k}+Q_k$. If $Q_k$ remains constant, $P_k$ grows polynomially. Although this growth is significantly slower than the exponential growth of $R_k^{-1}$ in CF-RLS, it still poses a problem if sustained for an extended period.

Additionally, another issue arises when the input is sparse. For accurate parameter estimation in such cases $Q_k$ needs to be large. However, this also means that the estimation results become highly sensitive to noise, leading to significant fluctuations.
\subsection{Proposed Algorithm: VDF-RLS}
Therefore, to address the issue of insufficient data richness over short time intervals, we apply the VDF-RLS algorithm. In this method, the forgetting factor is applied selectively, targeting only the portion of the information matrix that corresponds to the new input vector $u$. This involves first evaluating the singular vectors of the information matrix, where the singular values associated with these vectors quantify the amount of information in their respective directions. Zero or low singular values indicate a lack of information richness in those directions.  More specifically, we represent the  singular value decomposition of information matrix as:
\begin{align}\label{eqn:decouple_information}
    R_{K}=V_K\Sigma_KV_K^\top = \sum_i \sigma_{K}(i)v_K(i) v_K^\top(i),
\end{align}
where $\sigma_{K}(i)$ and $v_K(i)$ correspond to a $i$th singular value and singular vector of information matrix $R_K$ at  $K$th time instance. An input vector at the next instance $u_{K+1}$ can then be evaluated to determine how much it is aligned with $v_K(i)$; here, the alignment is evaluated by the inner product $u_{K+1}^\top v_K(i)$.  The forgetting factor for each direction is then ascribed as:
\begin{align}\label{eqn:VDF_forgetting}
    \lambda_i = \begin{cases}
\lambda\in(0,1) \quad\text{if }|v_K(i)^\top u_{K+1}|>\epsilon\\
1\quad\text{else}
\end{cases}
\end{align}

In this approach, the forgetting factor is applied only when the inner product between the new input vector and the singular vectors exceeds $\epsilon$, a certain threshold. By adjusting this threshold, we can enhance the algorithm's robustness against noise. Specifically, when the magnitude of the input vector is small, it is treated as noise; even if the angle between the input vector and the information direction is aligned, the forgetting factor is not applied in such cases.

With the forgetting factor defined in (\ref{eqn:VDF_forgetting}), the information matrix can be updated as follows:
\begin{align}\label{eqn:VDF_update_Rk}
\begin{split}
    R_{K+1} &=\sum_i \lambda_i\sigma_{K}(i)v_K(i) v^\top_K(i)+u_{K+1}u^\top_{K+1}.
\end{split}
\end{align}

Then, the new estimation can be obtained recursively using (\ref{eqn:CF-RLS}).

\section{Simulation results}
The proposed impedance estimation algorithm was implemented and tested using the MATLAB Simscape toolbox with a step size of 0.05 ms.

The simulation setup follows the structure shown in Fig. \ref{fig:circuits} (b),
where an inverter is connected to the grid with a voltage of $480,V_{\text{rms L-L}}$ and a frequency of 59.99 Hz, through a variable impedance. The inverter operates in grid-following mode using an average model, adjusting its output to meet a specified power setpoint.

\subsection{Performance comparison between different estimation algorithms}
In this section, the estimation performance of different algorithms is evaluated. The RLS algorithm with and without a forgetting factor, the Kalman filter-based algorithm, and the proposed VDF-RLS algorithm are compared. 

For the VDF-RLS algorithm, just as before, $\lambda = 0.995$ and $\epsilon=0.2$ are used. Similarly, the CF-RLS algorithm employs a forgetting factor of $\lambda=0.99995$, as using $\lambda=0.995$ led to excessive fluctuations between excitations, 
making a direct comparison with the same $\lambda$ meaningless.
Lastly, the Kalman filter is set with $S_k = \lambda =0.995$ to ensure similarity, as discussed in the previous section, and $Q_K = 0.00001I_2$, which is tuned to initially exhibit a similar adaptation speed to the VDF-RLS algorithm.

\begin{figure}[h]
    \centering
    \includegraphics[width=1\columnwidth]{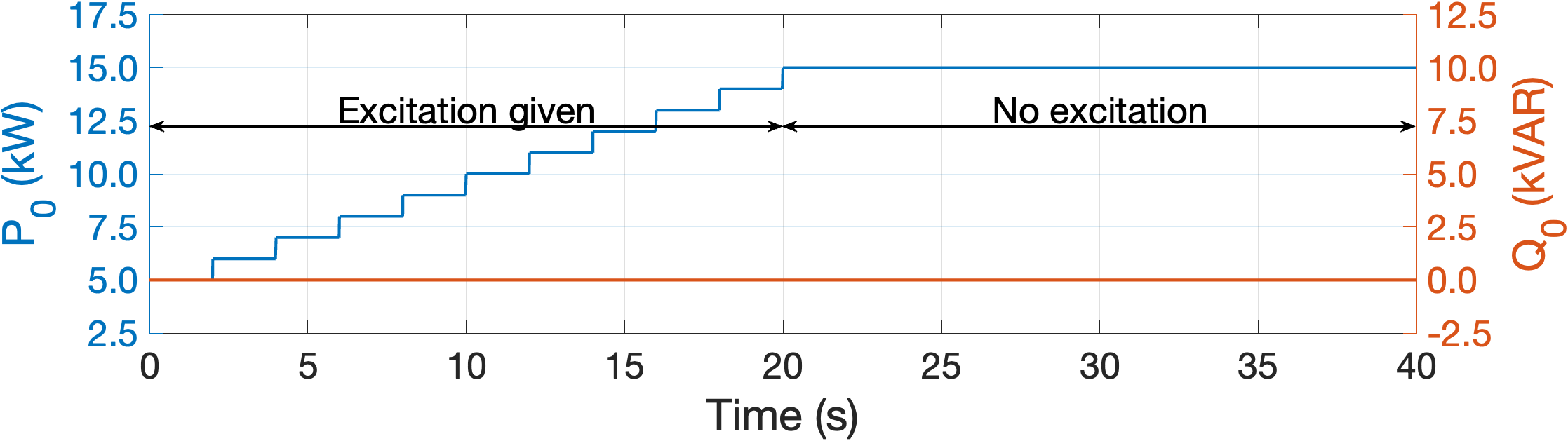}
    \caption{Power setpoints: Between $t=0$ and $t=20$ seconds, the setpoints change every 2 seconds, providing excitation. After $t=20$ seconds, no further excitation is applied.}
    \label{fig:PQ_setpoint_long}
\end{figure}

The initial conditions for the parameter estimation and the information matrix are set as $\widehat{\theta}_0=[0,0]^\top$ and $R_0=P_0^{-1}=0.001I_2$. Additionally, the PLL is designed to have a gain crossover frequency of $1$ Hz and for the band-pass filter used in preconditioning the data, the lower cutoff frequency is chosen to be $10$ Hz, and the upper cutoff frequency is $100$ Hz. Lastly, the power setpoints are provided as shown in Fig. \ref{fig:PQ_setpoint_long}, with white noise added to the measurements.

\begin{figure}[h]
    \centering
    \vspace{0.3cm}
    \begin{subfigure}[b]{1\columnwidth}
        \includegraphics[width=1\columnwidth]{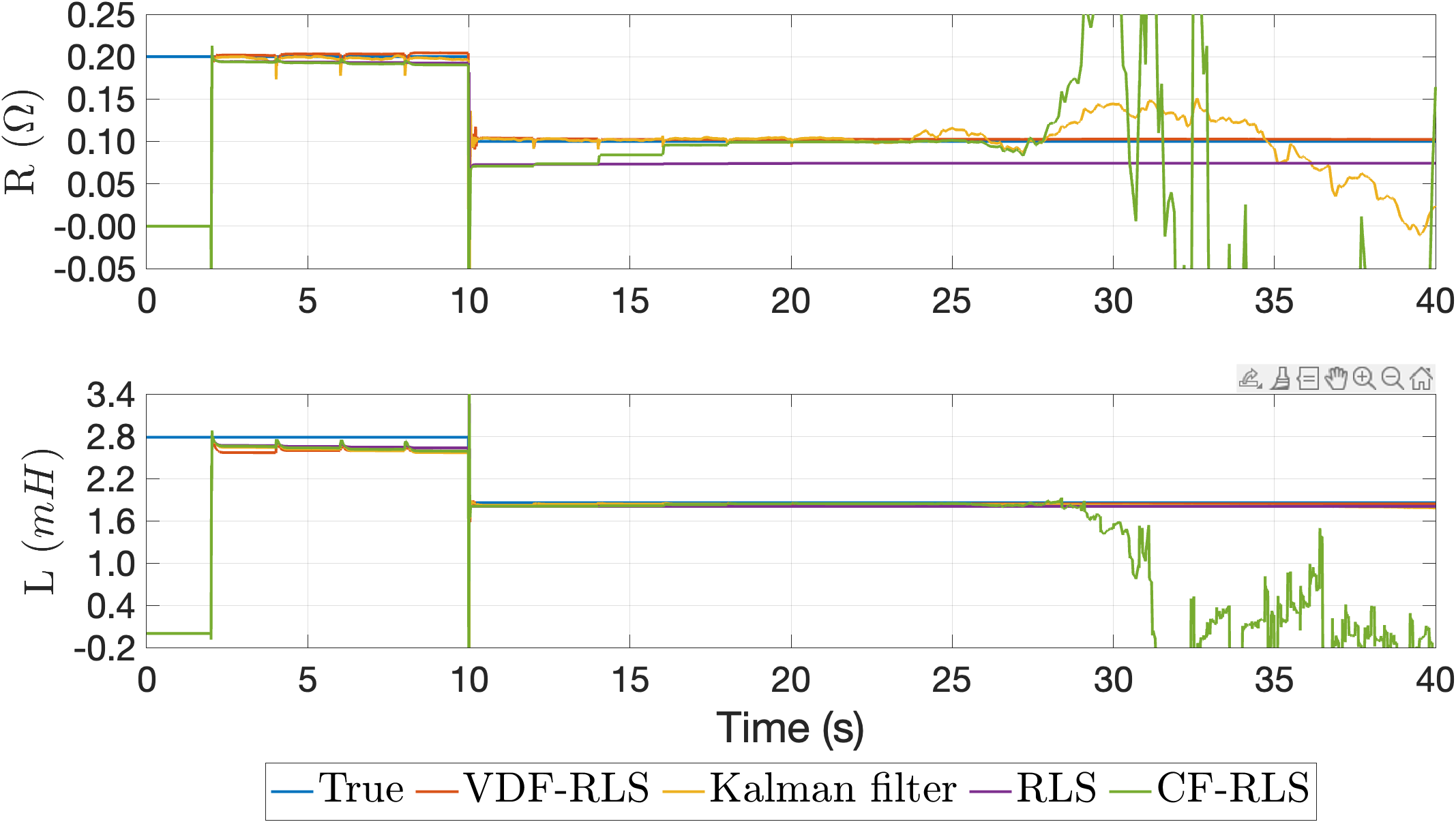}
        \caption{Impedance estimation results}
    \end{subfigure}\\
    \begin{subfigure}[b]{1\columnwidth}
        \includegraphics[width=1\columnwidth]{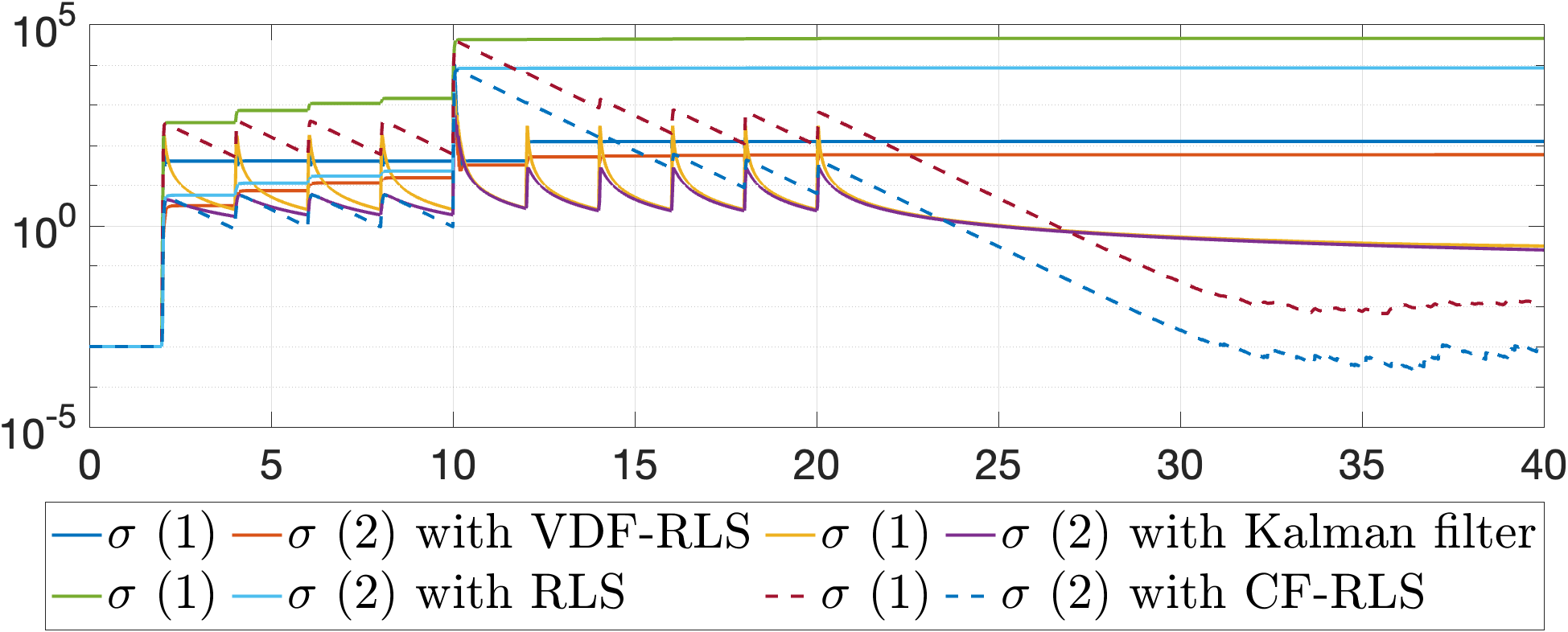}
        \caption{Singular values of information matrix}
    \end{subfigure}\\
    \caption{Line impedance estimation results using different algorithms. Impedance changes at $t=10$ seconds, while estimation begins at $t=2$ seconds to allow the PLL to lock onto the frequency.}
    \label{fig:WO_PE_noise}
\end{figure}

The simulation results are presented in Fig. \ref{fig:WO_PE_noise} and the estimation performance is evaluated by the Root Mean Square Percent Error (RMSPE) for each parameter (see Table \ref{tbl:RMSPE_algorithms}),  where 
\begin{align}
    \begin{split}\label{eqn:rmspe_def}
        RMSPE=\sqrt{\frac{1}{N}\sum_{k=1}^N\bigg(\frac{\widehat{\theta}_k(i)-\theta_k(i)}{\theta_k(i)}\bigg)^2} \times 100 \%.
    \end{split}
\end{align}
Here, due to a significant initial transient affecting the RMSPE, the evaluation is conducted using data from 15 to 20 seconds and 35 to 40 seconds.

The VDF-RLS algorithm demonstrates superior performance in both adapting to impedance changes and maintaining stable estimation under no-excitation conditions. In contrast, the RLS algorithm without forgetting fails to adapt to impedance changes. While the Kalman filter and CF-RLS successfully adjust to impedance changes when excitation is present, they become highly sensitive to measurement noise once excitation ceases, leading to significantly higher RMSPE values. Specifically, as seen in Table \ref{tbl:RMSPE_algorithms} for the without excitation case, for resistance estimation, the Kalman filter and CF-RLS achieve 60.3797\% and 785.3611\%, respectively—approximately 24 times and 321 times higher than that of VDF-RLS (2.4459\%). For inductance estimation, they reach 3.1674\% and 104.9205\%, which are 3 times and 103 times greater than the VDF-RLS result (1.016\%).

{\em Remark:} Overall, when excitation is present, the VDF-RLS, Kalman filter, and CF-RLS algorithms exhibit similar RMSPE, with any differences  likely due to tuning issues.
In contrast, when there is no more excitation (after $t=20$s), the Kalman filter and CF-RLS algorithms fail to maintain their estimation accuracy, becoming highly susceptible to measurement noise fluctuations. This degradation occurs due to the loss of information, as shown in Fig. \ref{fig:WO_PE_noise} (b). One can tune the CF-RLS and Kalman filter to resist the effects of the no-excitation condition for a longer duration. However, this merely slows down the rate at which the information matrix is discounted. Given enough time, these algorithms will inevitably become susceptible to noise fluctuations, leading to instability in the estimation.

\begin{table}[h]
\caption{RMSPE of R and L estimation under different algorithms}
\label{tbl:RMSPE_algorithms}
\centering
\setlength\tabcolsep{2pt}
    \begin{tabular}{|c|l|c|c|}
        \hline
        & \textbf{Algorithm used} & \textbf{\shortstack{RMSPE R(\%)}} & \textbf{\shortstack{RMSPE L(\%)}} \\ \hline
        \multirow{4}{*}{\shortstack{With excitation \\ Measured from\\$t = 15$ to $t = 20$s}} 
        & VDF-RLS(Proposed)      & 2.3426&1.1032 \\ \cline{2-4} 
        & Kalman filter         & 3.1979&1.1169 \\ \cline{2-4} 
        & RLS w/o forgetting & 26.315&2.8145 \\ \cline{2-4} 
        & CF-RLS                & 7.7352&1.5784 \\ \hline
        \multirow{4}{*}{\shortstack{Without excitation \\ Measured from\\$t = 35$ to $t = 40$s}} 
        & VDF-RLS(Proposed)      & 2.4459&1.016 \\ \cline{2-4} 
        & Kalman filter         & 60.3797&3.1674 \\ \cline{2-4} 
        & RLS w/o forgetting & 25.7157&2.8077 \\ \cline{2-4} 
        & CF-RLS                & 785.3611&104.9205 \\ \hline
    \end{tabular}
\end{table}

\subsection{Evaluation on preconditioning}
The proposed preconditioning consists of three main steps: (i) discarding q-axis dynamics and using only d-axis dynamics, (ii) employing a lower-bandwidth PLL to generate the rotating frame, and (iii) applying a bandpass filter. In this section, each method will be evaluated separately.

During the evaluation, the power setpoint changes in Fig. \ref{fig:PQ_setpoint_long}, between $t=0$ and $t=20$s are used, and white noise is added to the measurements as in previous tests. Additionally, throughout the assessment of the preconditioning effect, the VDF-RLS algorithm and the bandpass filter retain the same parameter settings as in previous evaluations.

\subsubsection{Comparison between using only d axis dynamics and using both d,q axis dynamics}
First, we compare the case when both the d- and q-axis dynamics are used. Due to the change in dimensions, we modify the input and output as follows: $u=\begin{bmatrix}BPF(i^d)&\big(BPF(d{i^d}/dt)-\omega BPF(i^q)\big)/\omega_0\\
BPF(i^q)&\big(BPF(d{i^q}/dt)+\omega BPF(i^d)\big)/\omega_0\end{bmatrix}$ and $y=\begin{bmatrix} BPF(V_c^d)\\ BPF(V_c^q)    
\end{bmatrix}$.
Additionally, during the update process, we evaluate the alignment of the input vector using $\|v_K(i)^\top u_{K+1}\|_2$ instead of the absolute value to accommodate the dimensional changes.

\begin{figure}[h]
    \centering
    \begin{subfigure}[b]{0.5\columnwidth}
    \includegraphics[width=\textwidth]{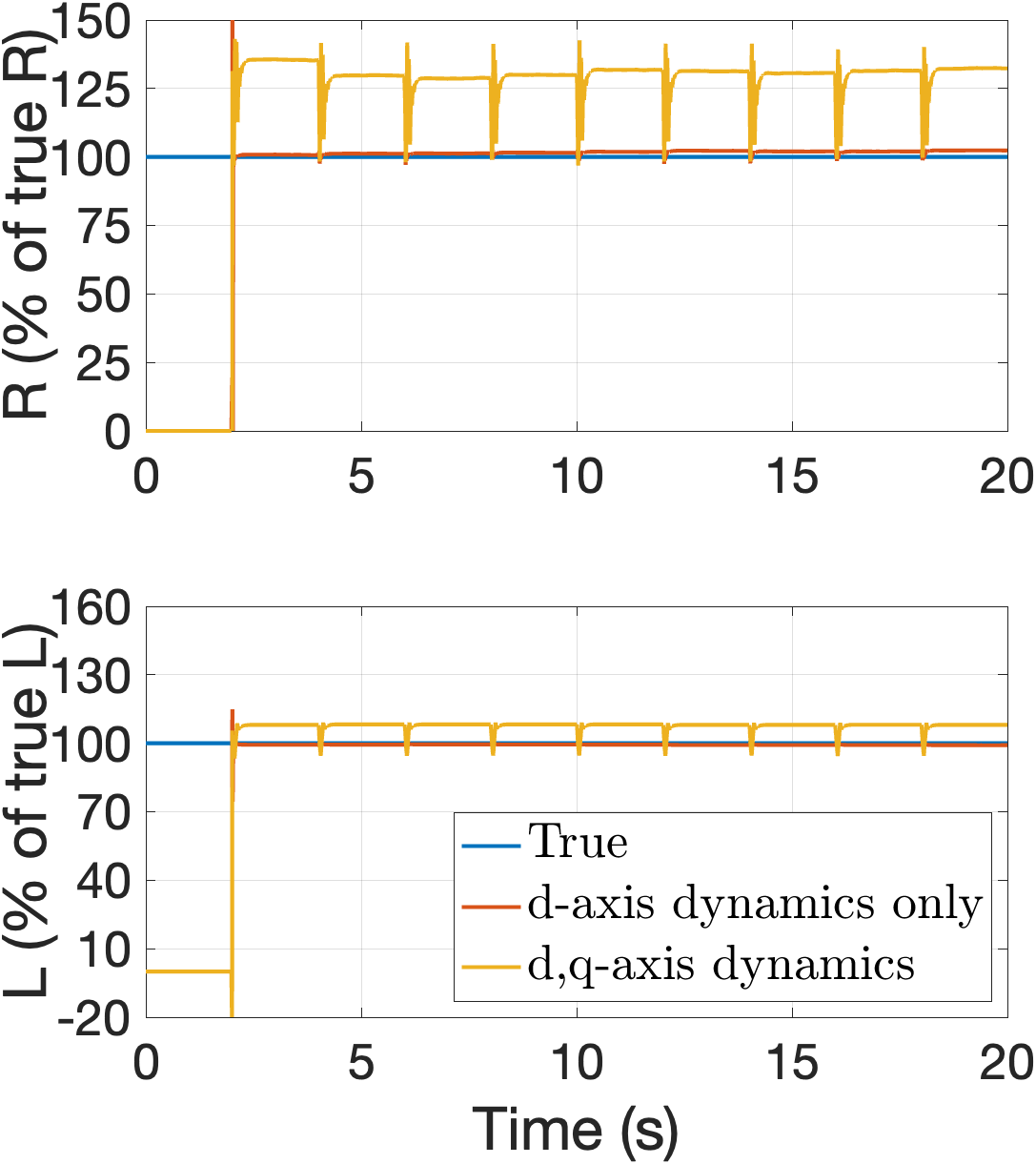}
    \caption{}
    \end{subfigure}%
    \begin{subfigure}[b]{0.5\columnwidth}
    \includegraphics[width=\textwidth]{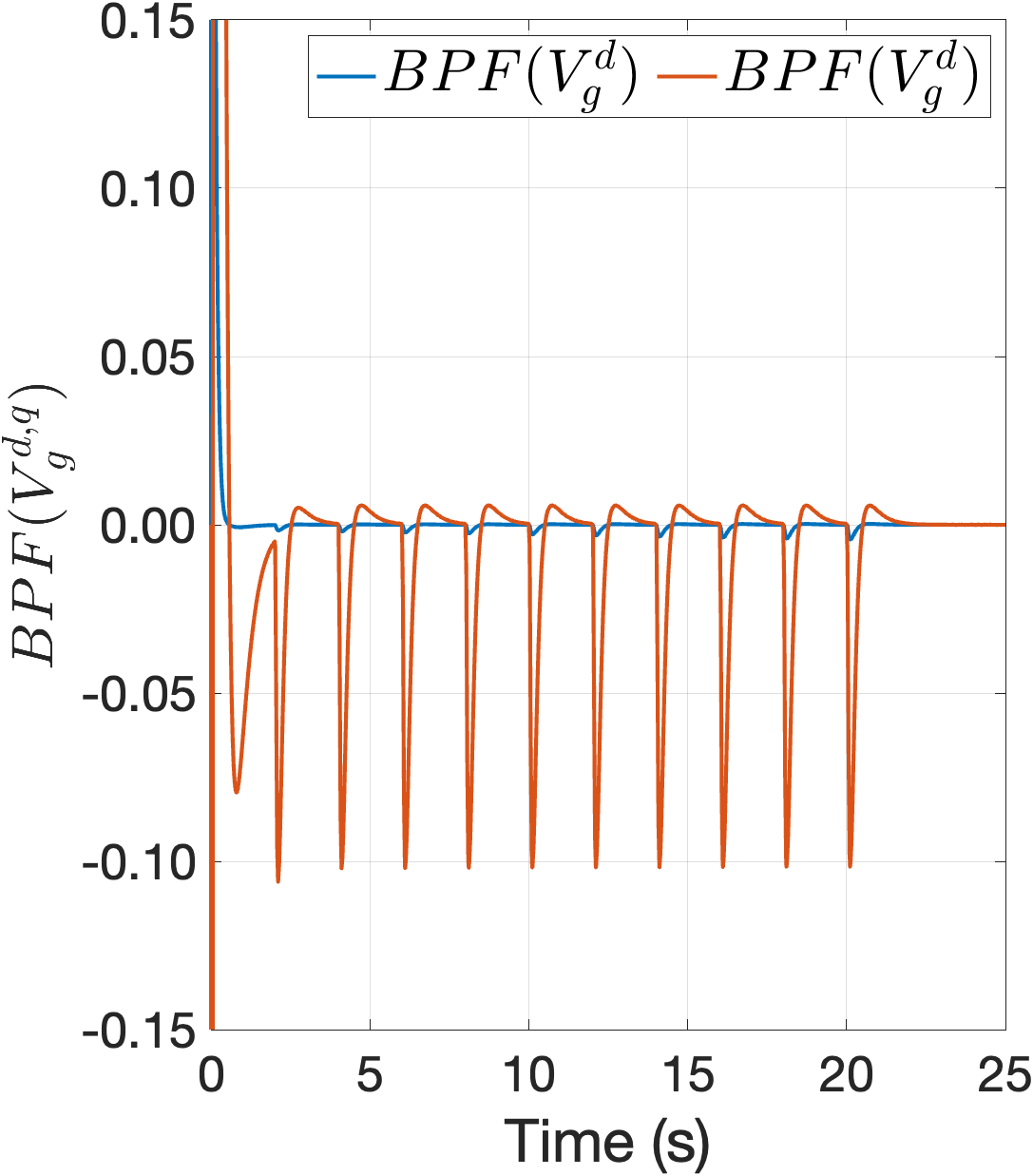}
    \caption{}
    \end{subfigure}\\
    \caption{(a) Estimation result comparison between using only d axis and both d,q axis dynamics (b) noise related term ($BPF(V_g^{d,q})$)}
    \label{fig:dq_dynamics_estimation_comparison}
\end{figure}

As shown in Fig. \ref{fig:dq_dynamics_estimation_comparison} (a), when only $d$-axis dynamics are used, the RMSPE values are much lower compared to the case where both $d$ and $q$-axis dynamics are used, despite the additional information provided by the $q$-axis dynamics. Specifically, the RMSPE for resistance is 2.1261\%, and for inductance, it is 0.78482\%—approximately 1/14 and 1/10 of the values observed when both $d$ and $q$-axis dynamics are used.

This fluctuation arises from noise. When using both $d$ and $q$ axis dynamics, the noise is given by $w=\begin{bmatrix} BPF(V_g^d)\\BPF(V_g^q)\end{bmatrix}$. Even within the same reference frame, $V_g^q$ experiences larger fluctuation , leading to larger $BPF(V_g^q)$ compare to $V_g^d$ and $BPF(V_g^d)$ as shown Fig. \ref{fig:dq_dynamics_estimation_comparison} (b). This is due to its larger sensitivity as discussed in previous section. In contrast, using only $d$ axis dynamics allows for effective rejection of this noise.

\subsubsection{Comparison between rotating frame frequency source}
In the previous section, we discussed the effect of the frequency source. To verify this, consider the following frequency sources:
\begin{itemize}
    \item Proposed low bandwidth PLL.\\
    A newly designed PLL with a gain crossover frequency of 1 Hz is implemented.
    \item PLL with high bandwidth - used in inverter control.\\
    The inverter's primary PLL, with a gain crossover frequency of 20 Hz, is utilized. This PLL ensures that $\psi_{inv} \approx 0$ at all times. However, phase changes induced by the inverter operation significantly affect the estimation results.
    \item Low pass filtered version of high bandwidth PLL.\\
    Instead of redesigning the PLL, a low-pass filter is applied to reduce the bandwidth of $\omega$. While this approach keeps $\delta\psi_{grid}$ in (\ref{eqn:vgdq_perturbation}) small, it does not guarantee accurate tracking of $\psi_{inv} \approx 0$. Two initial conditions are compared: $\psi_{inv}(0) = 0$ and $\psi_{inv}(0) = -\pi/6$. In this study, a low-pass filter with a cutoff frequency of 1 Hz is used.
\end{itemize}

\begin{figure}[h]
    \centering
    \includegraphics[width=1\columnwidth]{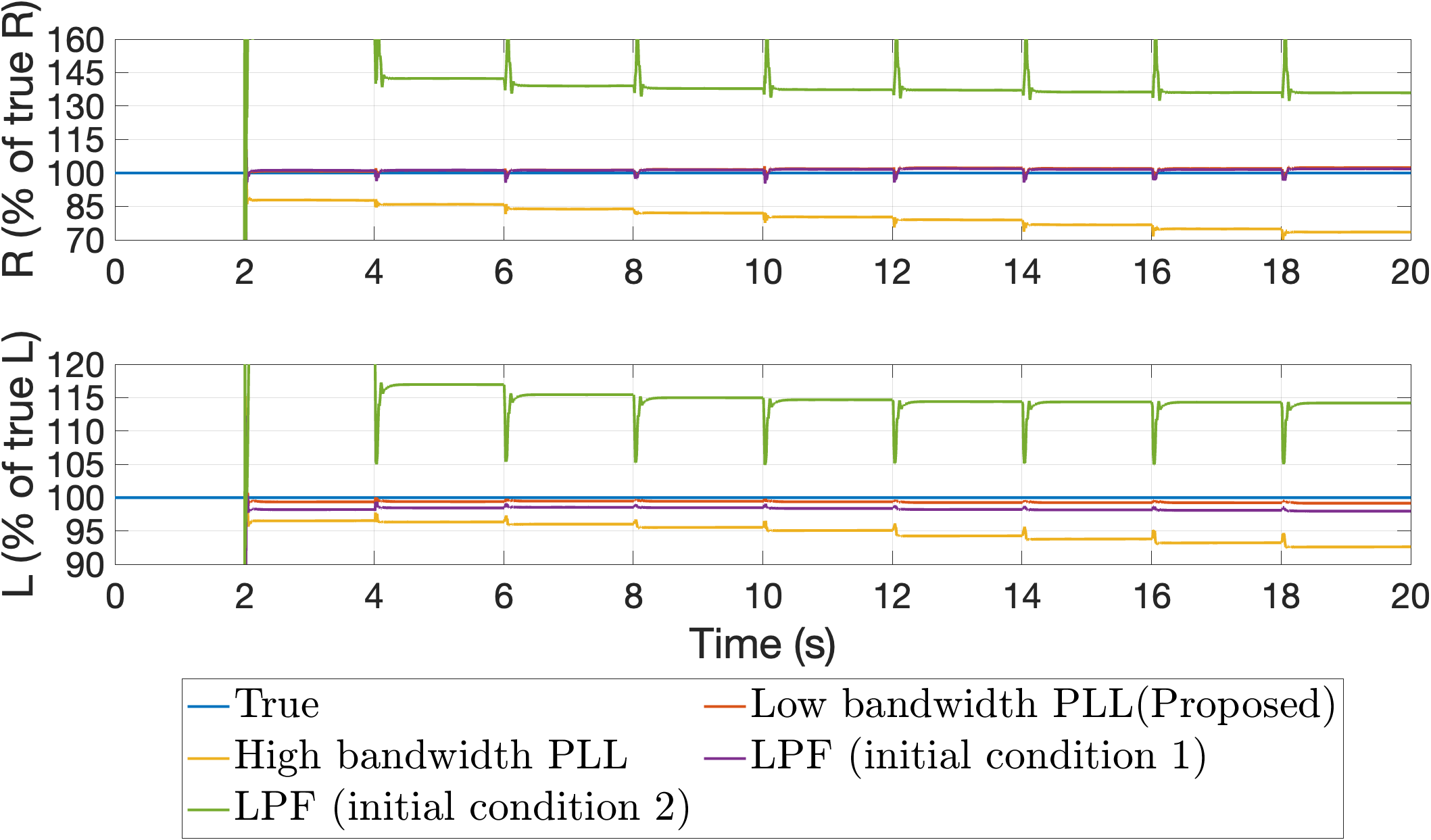}  
    \caption{Estimation results using different frequency as a source of rotating frame}
    \label{fig:freq_source_comparison}
\end{figure}

\begin{table}[h]
\caption{RMSPE of R and L estimation under different preconditioning methods (measured from $t=15$ to $t=20$s).}
\label{tbl:RMSPE}
\begin{center}
\setlength\tabcolsep{0.5pt}
\begin{tabular}{|c|c|c|}
\hline
\textbf{Preconditioning method} &  \textbf{\shortstack{RMSPE R(\%)}} & \textbf{\shortstack{RMSPE L(\%)}}\\
\hline
Proposed preconditioning method& 2.1261&0.78482\\
\hline
\shortstack{Using both $d$ and $q$ dynamics}&30.8127&7.8948\\
\hline
\shortstack{Using Primary (High bandwidth) PLL freq\\ as a source of rotating frame}&25.3133&6.902\\
\hline
\shortstack{Using low pass filtered primary PLL freq\\ as a source of rotating frame\\ (with initial condition $\psi_{inv}(0)=0$)}&1.689&1.933\\
\hline
\shortstack{Using low pass filtered primary PLL freq\\ as a source of rotating frame\\ (with initial condition $\psi_{inv}(0)=-\frac{\pi}{6}$)}&36.7264&14.0925\\
\hline
\shortstack{Filtering the signal using\\Derivative + Low pass filter}&5.6093&9.7878\\
\hline
\end{tabular}
\end{center}
\end{table}

As seen in Fig. \ref{fig:freq_source_comparison} and Table \ref{tbl:RMSPE}, the high-bandwidth PLL results in much higher RMSPE values—25.3133\% for resistance estimation and 6.902\% for inductance estimation—compared to the proposed method, where a low-bandwidth PLL is used.

The low-pass filtered frequency source, while performing better than the high-bandwidth PLL in some cases, shows significant variation in RMSPE depending on the $\psi_{inv}$ value. In one condition, it achieves RMSPE values of 1.689\% for resistance and 1.933\% for inductance, but in another, the RMSPE increases dramatically to 36.7264\% and 14.0925\%. This inconsistency makes the algorithm unreliable.

\subsubsection{Applying bandpass filter}
Lastly, we propose applying a bandpass filter to separate $V_g^{d,q}$ from the rest of the dynamics, whereas other approaches rely on the difference between two consecutive time instances, which is equivalent to taking a discrete-time derivative. Also, because many approaches implicitly use low-pass filtered data to mitigate measurement noise, we consider such algorithm as a low pass filter with a derivative.

\begin{figure}[h]
    \centering
    \begin{subfigure}[b]{0.5\columnwidth}
    \includegraphics[width=\textwidth]{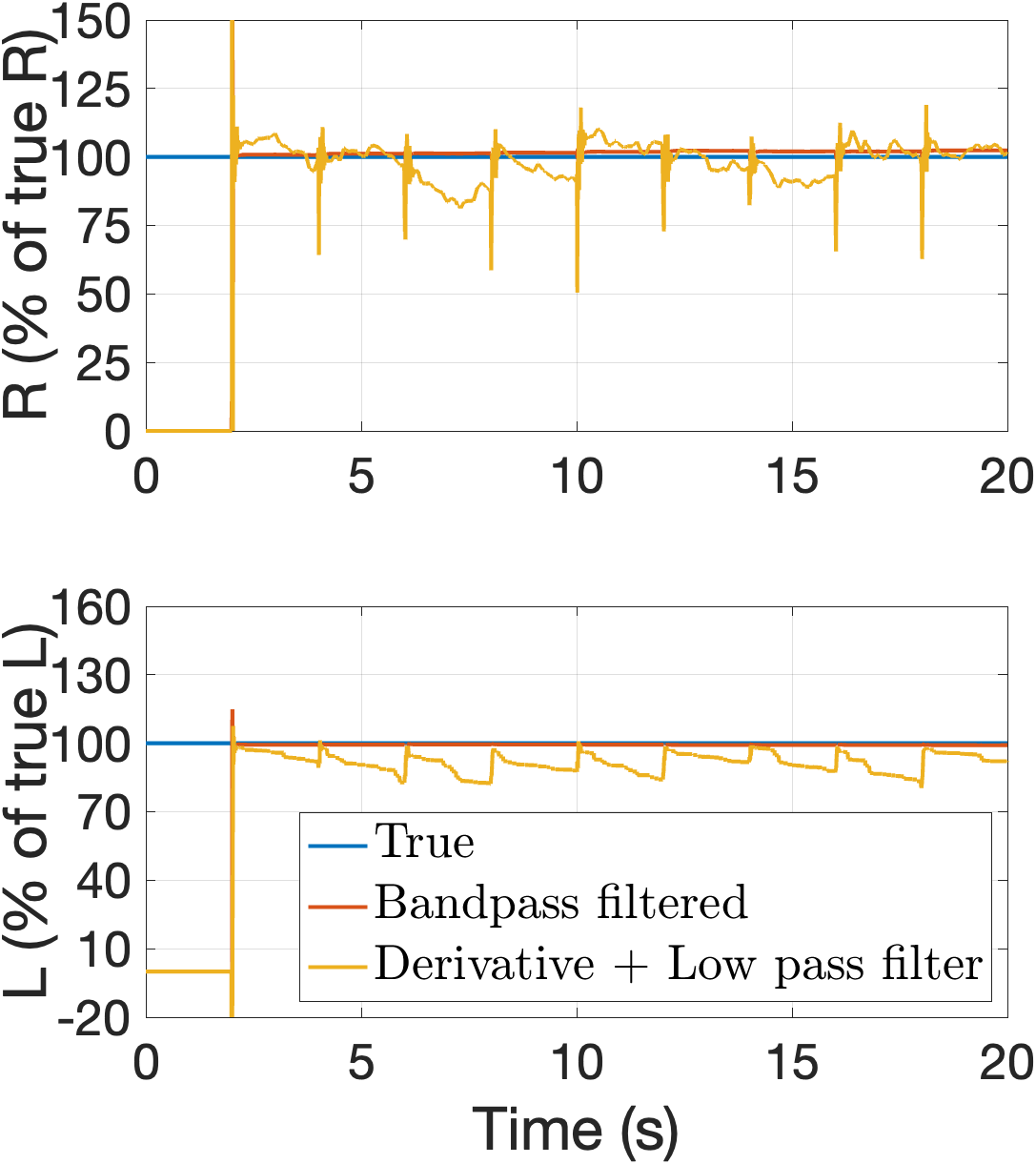}
    \caption{}
    \end{subfigure}%
    \begin{subfigure}[b]{0.5\columnwidth}
    \includegraphics[width=\textwidth]{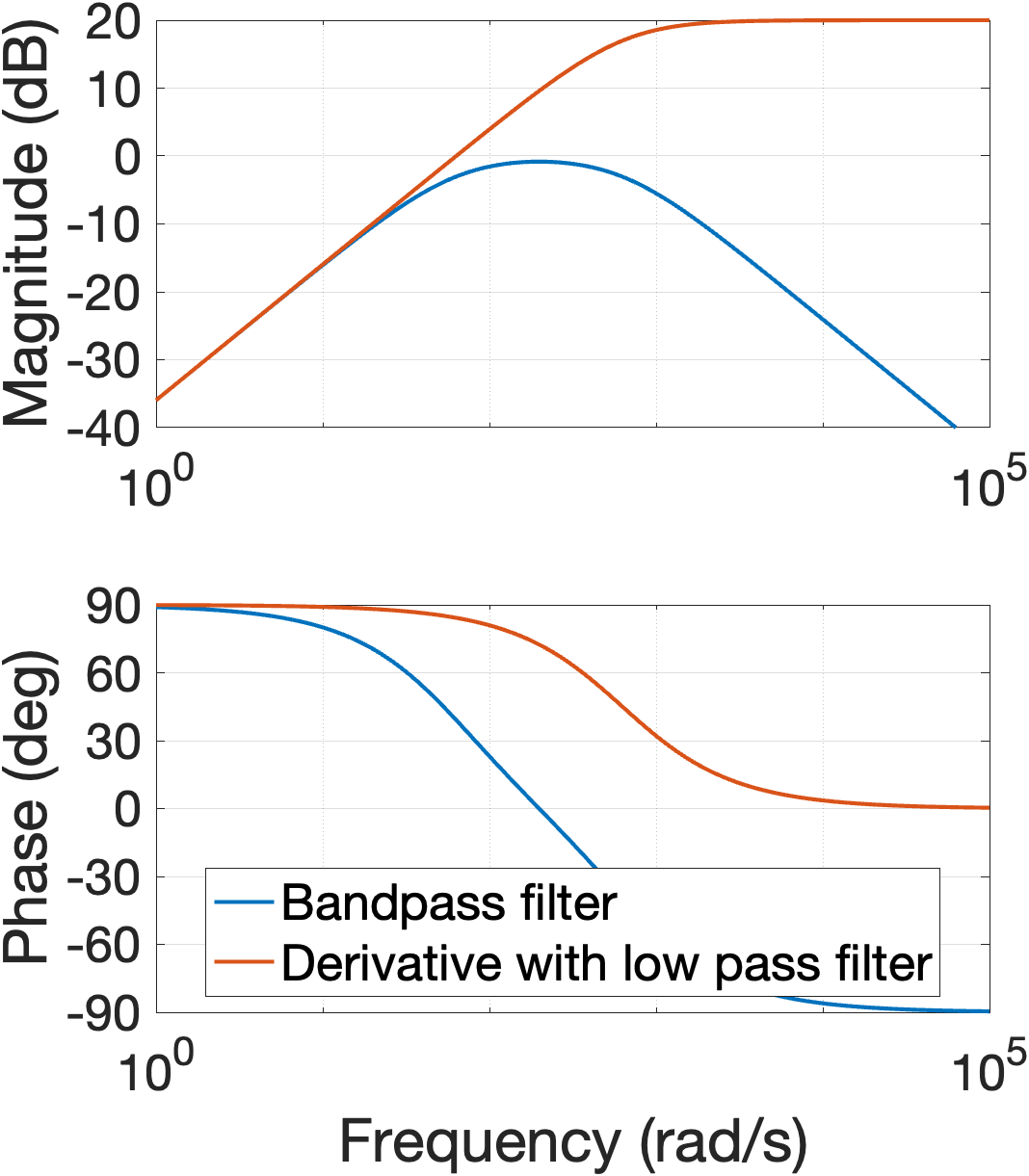}
    \caption{}
    \end{subfigure}\\
    \caption{(a) Estimation result using different preconditioning filter (b) Frequency response of different preconditioning filter}
    \label{fig:filter_comparison}
\end{figure}

Therefore, here we are comparing the following two filters
\begin{itemize}
    \item Proposed bandpass filter\\
    Filter with $\left(\frac{\omega_1}{s+\omega_1}\right)\left(\frac{s}{s+\omega_2}\right)$, where $\omega_1=200\pi$ rad/s (100Hz), and $\omega_2=20\pi$ rad/s (10Hz) is used as before
    \item Derivative with low pass filter\\
    Filter with $\left(\frac{\omega_1}{s+\omega_1}\right)\left(\frac{s}{\omega_2}\right)$ is used with the same $\omega_1$ and $\omega_2$ as above scenario. Here, $\frac{1}{\omega_2}$ is introduced to scale the response so that both test cases exhibit the same magnitude at low frequencies, as shown in Fig. \ref{fig:filter_comparison} (b).
\end{itemize}

Here, as seen in Fig. \ref{fig:filter_comparison} (a) and Table \ref{tbl:RMSPE}, the derivative with a low-pass filter is more vulnerable to measurement noise, resulting in RMSPE values of 5.6093\% and 9.7878\% for resistance and inductance estimation, respectively. In contrast, using a bandpass filter yields significantly lower RMSPE values of 2.1261\% and 0.78482\%. This is because the frequency response of the derivative does not roll off at high frequencies, preventing the preconditioning filter from effectively rejecting high-frequency noise, even with the low-pass filter, as shown in Fig. \ref{fig:filter_comparison} (b).

\section{CONCLUSIONS}

In this paper, we propose a VDF-RLS-based grid line parameter estimation algorithm with a preconditioning method. The VDF-RLS algorithm selectively updates historical information when new data is acquired, enabling fast estimation similar to CF-RLS and Kalman filters when excitation is present, while maintaining stability in the absence of persistent excitation, like the RLS algorithm without forgetting.

Additionally, the proposed preconditioning method leverages the dynamics of the grid impedance, reducing sensitivity to noise caused by inverter setpoint changes and measurement noise.

We validate this approach through simulations, comparing parameter estimation results using CF-RLS, Kalman filter, and RLS without forgetting algorithms. Different preconditioning methods are also evaluated in a similar manner.

\addtolength{\textheight}{-12cm}   

\bibliographystyle{Bibliography/IEEEtran}
\bibliography{Bibliography/CDC_2025} 

\end{document}